# Optimal Algorithms for Crawling a Hidden Database in the Web[*]

Cheng Sheng[1]    Nan Zhang[3]    Yufei Tao[1,2]    Xin Jin[3]

[1]Chinese University of Hong Kong    [2]Korea Advanced Institute of Science and Technology
{csheng, taoyf}@cse.cuhk.edu.hk              taoyf@kaist.ac.kr
[3]George Washington University
nzhang10@gwu.edu, xjin@gwmail.gwu.edu

## ABSTRACT

A *hidden database* refers to a dataset that an organization makes accessible on the web by allowing users to issue queries through a search interface. In other words, data acquisition from such a source is not by following static hyper-links. Instead, data are obtained by querying the interface, and reading the result page dynamically generated. This, with other facts such as the interface may answer a query only partially, has prevented hidden databases from being crawled effectively by existing search engines.

This paper remedies the problem by giving algorithms to extract all the tuples from a hidden database. Our algorithms are provably efficient, namely, they accomplish the task by performing only a small number of queries, even in the worst case. We also establish theoretical results indicating that these algorithms are asymptotically optimal – i.e., it is impossible to improve their efficiency by more than a constant factor. The derivation of our upper and lower bound results reveals significant insight into the characteristics of the underlying problem. Extensive experiments confirm the proposed techniques work very well on all the real datasets examined.

## 1. INTRODUCTION

It is known that existing search engines can reach only a small portion of the Internet. They crawl HTML pages interconnected with hyper-links, which constitute the so-called *surface web*. Nowadays, an increasing number of organizations (e.g., companies, governments, institutions, etc.) bring their data online, by allowing a public user to query their back-end databases through context-dependent web interfaces. More specifically, data acquisition is performed by *interacting* with the interface at runtime, as opposed to following hyper-links. As a result, those back-end databases cannot be effectively crawled by a search engine under the current technology, and therefore, are usually referred to as *hidden databases*.

[*]This work was supported in part by (i) US NSF grants 0915834 and 1117297, (ii) projects GRF 4169/09, 4166/10, 4165/11 from HKRGC, and (iii) the WCU (World Class University) program under the National Research Foundation of Korea, and funded by the Ministry of Education, Science and Technology of Korea (Project No: R31-30007).

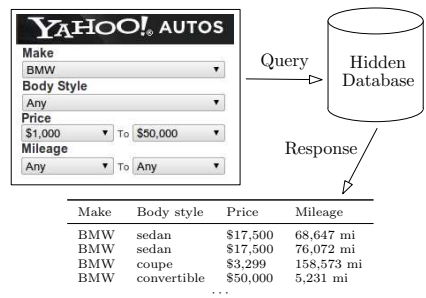

**Figure 1: Form-based querying of a hidden database**

Consider, for example, *Yahoo! Autos* (*autos.yahoo.com*), a popular website for online trading of automobiles. A potential buyer specifies her/his filtering criteria through a form as illustrated in Figure 1. The query is submitted to the system, which runs it against the back-end database, and returns the result to the user. What makes it non-trivial (for a search engine) to crawl the database is that, setting all search criteria to ANY does not accomplish the task. The reason is that a system typically limits the number $k$ of tuples returned ($k = 1000$ for *Yahoo! Autos*, at the time this paper was written), and that repeating the same query may not retrieve new tuples, i.e., the same $k$ tuples may always be returned.

The ability of crawling a hidden database comes with the appealing promise of enabling virtually any form of processing on the database's content. The challenge, however, is clear: how to obtain *all* the tuples, given that the system limits the number of return tuples for each query? A naive solution is to issue a query for every single location in the data space (e.g., in Figure 1, the data space is the Cartesian product[1] of the domains of MAKE, BODY STYLE, PRICE, and MILEAGE), but the number of queries needed can obviously be prohibitive. This gives rise to an interesting problem, as we define in the next subsection, where the objective is to *minimize* the number of queries.

### 1.1 Problem Definitions

We consider that the *data space* $\mathbb{D}$ has $d$ attributes $A_1, ..., A_d$, each of which has a discrete domain. Specifically, denote by $dom(A_i)$ the domain of $A_i$ for each $i \in [1, d]$; then, $\mathbb{D}$ is the Cartesian product of $dom(A_1), ..., dom(A_d)$. We refer to each element of the Cartesian product as a *point* in $\mathbb{D}$, i.e., a point is a possible combination of values of all dimensions.

---
[1]While one may leverage knowledge of attribute dependencies - e.g., BMW does not sell trucks in the US - to prune the data space into a subset of the Cartesian product, the subset is often still too large to enumerate.

Depending on whether there is a total ordering on $dom(A_i)$ or not, we call $A_i$ a *numeric* or *categorical* attribute, respectively. Our discussion distinguishes three types of $\mathbb{D}$:

- **Numeric:** all the $d$ attributes of $\mathbb{D}$ are numeric.
- **Categorical:** all the $d$ attributes are categorical. In this case, we use $U_i$ to represent the size of $dom(A_i)$, i.e., how many distinct values there are in $dom(A_i)$.
- **Mixed:** the first $cat \in [1, d-1]$ attributes $A_1, ..., A_{cat}$ are categorical, whereas the other $d - cat$ attributes are numeric. Similar to before, let $U_i = |dom(A_i)|$ for each $i \in [1, cat]$.

To facilitate presentation, we consider that the domain of a numeric $A_i$ to be the set of all integers, whereas that of a categorical $A_i$ to be the set of integers from 1 to $U_i$. Keep in mind, however, that the ordering of these values is *irrelevant* to a categorical $A_i$.

Let $D$ be the hidden database of a server with each element of $D$ being a point in $\mathbb{D}$. To avoid ambiguity, we will always refer to elements of $D$ as *tuples*. $D$ is a *bag* (i.e., a multi-set), that is, it may contain identical tuples.

The server supports *queries* on $D$. As shown in Figure 1, each query specifies a predicate on each attribute. Specifically, if $A_i$ is numeric, the predicate is a range condition in the form of

$$A_i \in [x, y]$$

where $[x, y]$ is an interval in $dom(A_i)$. On the other hand, for a categorical $A_i$, the predicate is:

$$A_i = x$$

where $x$ is either a value in $dom(A_i)$ or a wildcard $\star$. In particular, a predicate $A_i = \star$ means that $A_i$ can be an arbitrary value in $dom(A_i)$, i.e., capturing BODY STYLE = ANY in Figure 1. Note that if a hidden database server only allows single-value predicates (i.e., no range-condition support) on a numeric attribute, then we can simply consider the attribute as categorical.

Given a query $q$, denote by $q(D)$ the bag of tuples in $D$ qualifying all the predicates of $q$. The server does not necessarily return the entire $q(D)$ – it does so *only* when $q(D)$ is small. Formally, the response of the server is:

- if $|q(D)| \leq k$: the entire $q(D)$ is returned. In this case, we say that $q$ is *resolved*.
- Otherwise: only $k$ tuples[2] in $q(D)$ are returned, together with a signal indicating that $q(D)$ still has other tuples. In this case, we say that $q$ *overflows*.

The value of $k$ is a system parameter (e.g., $k = 1000$ for *Yahoo! Autos*, as mentioned earlier). It is important to note that, in case a query $q$ overflows, repeatedly issuing the same $q$ may always get the same response from the server, and does not help to obtain the other tuples in $q(D)$.

The problem addressed by this paper is:

PROBLEM 1. *(HIDDEN DATABASE CRAWLING) Retrieve the entire $D$ while minimizing the number of queries.*

Recall that $D$ is a bag, i.e., it may have duplicate tuples. We require that no point in the data space $\mathbb{D}$ have *more than $k$* tuples in $D$. Otherwise, Problem 1 has no solution at all. To see this, consider the existence of $k+1$ tuples $t_1, ..., t_{k+1}$ in $D$, all of which are equivalent to a point $p \in \mathbb{D}$. Then, whenever $p$ satisfies a query, the server can *always* choose to leave $t_{k+1}$ out of its response, making it impossible for any algorithm to extract the entire $D$. Note that, in *Yahoo! Autos*, the previous requirement essentially states that there cannot be $k = 1000$ vehicles in the database having exactly the same values on *all* attributes – an assumption that is fairly realistic.

---
[2]In practice, these are usually the $k$ tuples that have the highest priorities (e.g., according to a ranking function) among all the tuples qualifying the query.

As mentioned in Problem 1, the *cost* of an algorithm is the number of queries issued. This metric is motivated by the fact that, most systems have a control on how many queries can be submitted by the same IP address within a period of time (e.g., a day). Therefore, a crawler must minimize the number of queries to get the task done, besides bringing the burden of the server to the lowest level.

We will use $n$ to denote the number of tuples in $D$. It is clear that the number of queries needed to extract the entire $D$ is at least $n/k$. Of course, this ideal cost may not always be possible. Hence, the central (technical) questions to be answered are two-fold. First, on the upper bound side, how to solve Problem 1 by performing only a small number of queries even in the *worst* case? Second, on the lower bound side, how many queries are *compulsory* for solving the problem in the worst case?

## 1.2 Our Results

This paper presents a systematic study of *hidden database crawling* as defined in Problem 1. At a high level, our first contribution is a set of algorithms that are both provably fast in the worst case, and efficient on practical data. Our second contribution is a set of lower-bound results establishing the *hardness* of the problem. These results make explicit how the hardness is affected by the underlying factors, and thus reveal valuable insights into the characteristics of the problem. Furthermore, the lower bounds also prove that our algorithms are already optimal asymptotically, i.e., they cannot be improved by more than a constant factor.

Our first main result is:

THEOREM 1. *There is an algorithm for solving Problem 1 whose cost is:*

- $O(d \cdot \frac{n}{k})$ *when $\mathbb{D}$ is numeric;*
- *at most $U_1$ when $\mathbb{D}$ is categorical and $cat = 1$ (i.e., there is only one categorical attribute);*
- *at most $\frac{n}{k} \cdot \sum_{i=1}^{d} \min\{U_i, \frac{n}{k}\} + \sum_{i=1}^{d} U_i$ when $\mathbb{D}$ is categorical and $cat > 1$;*
- *at most $U_1 + O(d \cdot \frac{n}{k})$ when $\mathbb{D}$ is mixed and $cat = 1$;*
- *otherwise (i.e., $\mathbb{D}$ is mixed and $cat > 1$): at most*

$$\frac{n}{k} \cdot \sum_{i=1}^{cat} \min\left\{U_i, \frac{n}{k}\right\} + \sum_{i=1}^{cat} U_i + O\left((d - cat)\frac{n}{k}\right).$$

The above can be conveniently understood as follows: our algorithm pays an (additive) cost of $O(n/k)$ for each numeric attribute $A_i$, whereas it pays $\frac{n}{k} \cdot \min\{U_i, \frac{n}{k}\} + U_i$ for each categorical $A_i$. The only exception is when $cat = 1$: in this scenario, we pay merely $U_1$ for the (only) categorical attribute $A_1$. Notice that the cost on each numeric attribute is *irrelevant* to its domain size.

Our second main result complements the preceding one:

THEOREM 2. *<u>None</u> of the results in Theorem 1 can be improved by more than a constant factor in the worst case.*

Besides establishing the optimality of our upper bounds in Theorem 1, Theorem 2 has its own interesting implications. First, it indicates the unfortunate fact that, for all types of $\mathbb{D}$, the best achievable query time (in the worst case) is much higher than the ideal cost of $n/k$ (nevertheless, Theorem 1 suggests that we can achieve this cost *asymptotically* when $d$ is a constant and all attributes are numeric). Second, as the number *cat* of categorical attributes increases from 1 to 2, the discrepancy of the time complexities in Theorem 1 is not an artifact, but rather, it is due to an inherent *leap* in the hardness of the problem (this is true regardless of the number of numeric attributes). That is, while we pay only $O(U_1)$ extra queries for the (sole) categorical attribute when $cat = 1$, as *cat* grows to 2 and onwards, the cost paid (by any algorithm) for each categorical $A_i$ has an extra term of $\frac{n}{k} \min\{U_i, \frac{n}{k}\}$. Given that the term is multiplicative, this finding implies (perhaps surprisingly) that, in the worst case, it may be infeasible to crawl a hidden database with a large size $n$, and at least 2 categorical attributes such that at least one of them has a large domain.

We have performed extensive experiments to evaluate the efficiency of the proposed algorithms on real datasets, and demonstrate that they demand significantly fewer queries than alternative solutions. Our experimentation also reveals that the number of queries needed to crawl a hidden database may be far less than previously thought; for example, for $k = 1000$, around 200 queries already suffice for crawling a dataset containing 69,768 tuples from the hidden database at *Yahoo! Autos*. This phenomenon suggests that, for a search engine, crawling a hidden database may no longer be a goal of tomorrow, whereas for a data provider, permitting an engine to crawl its database is not expected to impose a heavy toll on its workload.

## 1.3 Practical Remarks

**Domain values.** In our problem definition, the crawler should know the domains of the categorical attributes (note that this issue is irrelevant to numerical attributes, whose domains can always be considered to be $(-\infty, \infty)$). For some websites, the domains of all attributes are explicitly provided such that our algorithms can be applied *immediately*. For example, this is the case for *Yahoo! Autos*, where all the values of, say, MAKE can be seen from the pull-down menu of its query interface. For other websites, before using our technique, the crawler needs to first discover the domain values of categorical attributes. Domain discovery has been studied in [15], and can be accomplished with a number of effective algorithms.

**Attribute dependency.** As mentioned in the above discussions, because of attribute dependencies in a practical hidden database, not all the points in the data space $\mathbb{D}$ can contain a tuple. For example, with proper external knowledge of the dependency between MAKE and BODY STYLE, one does not need to explore points with MAKE = BMW and BODY STYLE = TRUCK. While knowledge of attribute dependencies is not considered in this paper, please note that our upper bound results *still hold* even in scenarios where attribute dependencies exist. In other words, even if our algorithm may issue some unnecessary queries, the number of queries needed can nonetheless still be limited under the claimed bounds, as is exactly the merit of Theorem 1. In practice, there is an obvious heuristic for adapting our algorithm to account for attribute dependencies: the crawler issues a query demanded by our algorithm *only if* the query covers at least one valid point in $\mathbb{D}$ (according to the crawler's dependency knowledge). The query cost can only go *down*, i.e., still guaranteed to be below our upper bounds.

Acquisition of knowledge about attribute dependencies requires dedicated efforts to analyze the hidden database at a particular website – efforts that obviously cannot be afforded by the crawler for *all* websites. The implication is thus that the crawler may not be aware of the latent attribute dependencies at many of the websites being crawled. This is where our lower bound results come into place: they indicate the ultimate worst-case efficiency that the crawler can possibly achieve in these environments: a piece of information vital for the crawler's design.

## 1.4 Previous Work

A significant body of research has been carried out on how to extract, integrate, and analyze data from the *deep web*, a general term referring to the entire collection of online information unreachable by search engines (including, but not limited to, hidden databases). As explained below, however, the issues that have been addressed by the existing work are all orthogonal to this paper.

Most relevant to our work are the previous studies on crawling hidden text-based [1, 5, 18, 20] and structured [2, 7, 16, 17, 19] databases. The focus of those studies is how to formulate queries to retrieve meaningful results. More specifically, the primary challenge in [1, 5, 18, 20], where the query interface is a keyword-based (Google-like) form, is to discover legitimate query keywords. On the other hand, the main objective in [2, 7, 16, 17, 19], where the query interface is an HTML-form like the one in Figure 1, is to expose combinations of input values suitable for filling in the form. In this paper, we are not concerned with mining effective queries, but instead, attack directly how to acquire a complete hidden database with the smallest number of queries (see Problem 1).

Also relevant is the literature of data analytics in deep web. In this vein, a main stream is to investigate how sampling can be deployed to perform, for example, content summary generation [8, 14], top-$k$ retrieval [6], aggregate estimation [9], measurement of various metrics of search engines [3, 4], and so on. The crawling techniques proposed in this paper aim at enabling a much broader class of applications (e.g., virtually any query on the database, as described earlier), which otherwise would not be possible if only a sample of the hidden database could be obtained.

It is worth mentioning that, there has been considerable research on other problems related to the deep web, which, however, are only remotely related to our work. While a complete survey is out of the scope of this paper, entry points for further reading can be found in [10, 21] on parsing and understanding web interfaces, in [13] on attribute mapping across different interfaces, and in [11, 12] for integrating the query interfaces of multiple hidden databases.

## 2. NUMERICAL ATTRIBUTES

This section will explain how to solve Problem 1 when the data space $\mathbb{D}$ is numeric. In Section 2.1, we first define some atomic operators, and present an algorithm that is intuitive, but has no attractive performance bounds. Then, in Sections 2.2 and 2.3, we present another algorithm to achieve the optimal performance.

### 2.1 Basic Operations and Baseline Algorithm

Recall that, in a numeric $\mathbb{D}$, the predicate of a query $q$ on each attribute is a range condition. Thus, $q$ can be regarded as a $d$-dimensional (axis-parallel) rectangle, such that its result $q(D)$ consists of the tuples of $D$ covered by that rectangle. If the predicate of $q$ on attribute $A_i$ ($i \in [1, d]$) is $A_i \in [x_1, x_2]$, we say that $[x_1, x_2]$ is the *extent* of the rectangle of $q$ along $A_i$. Henceforth, we may use symbol $q$ to refer to its rectangle also, when no ambiguity can be caused. Clearly, settling Problem 1 is equivalent to determining the entire $q(D)$ where $q$ is the rectangle covering the whole $\mathbb{D}$.

**Split.** A fundamental idea to extract all the tuples in $q(D)$ is to refine $q$ into a set $S$ of smaller rectangles, such that each rectangle $q' \in S$ can be resolved (i.e., $q'(D)$ has at most $k$ tuples). Note that this always happens as long as rectangle $q'$ is sufficiently small – in the extreme case, when $q'$ has degenerated into a point in $\mathbb{D}$, the query $q'$ is definitely resolved (otherwise, there would be at least $k+1$ tuples of $D$ at this point). Therefore, a basic operation in our algorithms for Problem 1 is *split*, as described next.

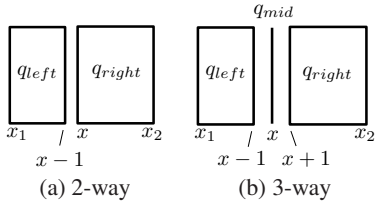

(a) 2-way  (b) 3-way

**Figure 2: Splitting**

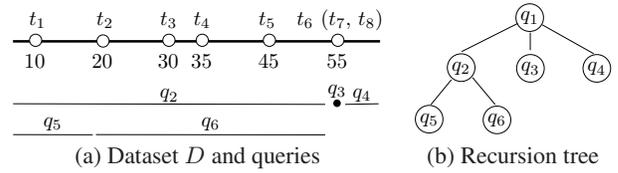

(a) Dataset $D$ and queries  (b) Recursion tree

**Figure 3: Illustration of 1d *rank-shrink***

Given a rectangle $q$, we may perform two types of splitting, depending on how many rectangles $q$ is divided into:

- **2-way split:** Let $[x_1, x_2]$ be the extent of $q$ on $A_i$ (for some $i \in [1, d]$). A *2-way split at a value* $x \in [x_1, x_2]$ partitions $q$ into rectangles $q_{left}$ and $q_{right}$, by dividing the $A_i$-extent of $q$ at $x$. Formally, on any attribute other than $A_i$, $q_{left}$ and $q_{right}$ have the same extents as $q$. Along $A_i$, however, the extent $q_{left}$ is $[x_1, x-1]$, whereas that of $q_{right}$ is $[x, x_2]$. Figure 2a illustrates the idea by splitting on the horizontal attribute.

- **3-way split:** Let $[x_1, x_2]$ be defined as above. A *3-way split at a value* $x \in [x_1, x_2]$ partitions $q$ into rectangles $q_{left}$, $q_{mid}$ and $q_{right}$ as follows. On any attribute other than $A_i$, they have the same extent as $q$. Along $A_i$, however, the extent of $q_{left}$ is $[x_1, x-1]$, that of $q_{mid}$ is $[x, x]$, and that of $q_{right}$ is $[x+1, x_2]$. See Figure 2b.

In the sequel, a 2-way split will be abbreviated simply as a split. No confusion can arise as long as we always mention 3-way in referring to a 3-way split. The extent of a query $q$ on an attribute $A_i$ can become so short that it covers only a *single* value, in which case we say that $A_i$ is *exhausted* on $q$. For instance, the horizontal attribute is exhausted on $q_{mid}$ in Figure 2b. It is easy to see that there is always a *non-exhausted* attribute on $q$ unless $q$ has degenerated into a point.

**Binary shrink.** Next, we describe a straightforward algorithm for solving Problem 1, which will serve as the baseline approach for comparison. This algorithm, named *binary-shrink*, repeatedly performs (2-way) splits until a query is resolved. Specifically, given a rectangle $q$, *binary-shrink* runs the rectangle (by submitting its corresponding query to the server) and finishes if $q$ is resolved. Otherwise, the algorithm splits $q$ on an attribute $A_i$ that has not been exhausted, by cutting the extent $[x_1, x_2]$ of $q$ along $A_i$ into equally long intervals (i.e., the split is performed at $x = \lceil (x_1 + x_2)/2 \rceil$). Let $q_{left}, q_{right}$ be the queries produced by the split. The algorithm then recurses on $q_{left}$ and $q_{right}$, respectively.

**Remark.** It is obvious that the cost of *binary-shrink* (i.e., the number of queries issued) depends on the domain sizes of the (numeric) attributes of $\mathbb{D}$, which can be *unbounded*. In the following subsections, we will improve this algorithm to optimality.

### 2.2 One-Dimensional Case

Before giving our ultimate algorithm for settling Problem 1 of any dimensionality $d$, in this subsection we first explain how it works for $d = 1$. This will clarify the rationale behind the algorithm's efficiency, and facilitate our analysis for a general $d$. It is worth mentioning that the presence of only one attribute removes the need to specify the split dimension in describing a split.

**Rank-shrink.** Our algorithm, named *rank-shrink*, differs from *binary-shrink* in two ways. First, when performing a (2-way) split, instead of cutting the extent of a query $q$ in half, we aim at ensuring that at least $k/4$ tuples fall in *each* of the rectangles generated by the split. Such a split, however, may not always be possible, which as we will see can happen if many tuples are identical to each other. Hence, the second difference that *rank-shrink* makes is to perform a 3-way split in such a scenario, which gives birth to a query (among the 3 created) that can be immediately resolved.

Formally, given a query $q$, the algorithm eventually returns $q(D)$. It starts by issuing $q$ to the server, which returns a bag $R$ of tuples. If $q$ is resolved, the algorithm terminates by reporting $R$. Otherwise (i.e., $q$ overflows), we sort the tuples of $R$ in ascending order, breaking ties arbitrarily. Let $o$ be the $(k/2)$-th tuple in the sorted order, with its $A_1$-value being $x$. Now, we count the number $c$ of tuples in $R$ identical to $o$ (i.e., $R$ has $c$ tuples with $A_1$-value $x$), and proceed as follows:

- **Case 1:** $c \le k/4$. Split $q$ at $x$ into $q_{left}$ and $q_{right}$, each of which must contain at least $k/4$ tuples in $R$. To see this for $q_{left}$ (symmetric reasoning applies to $q_{right}$), note there are at least $k/2 - c \ge k/4$ tuples of $R$ *strictly* smaller than $x$, all of which fall in $q_{left}$. The case for $q_{right}$ follows in analogy.

- **Case 2:** $c > k/4$. Perform a 3-way split on $q$ at $x$. Let $q_{left}$, $q_{mid}$ and $q_{right}$ be the resulting rectangles (note that, the ordering among them *matters*; see Section 2.2). Observe that $q_{mid}$ has degenerated into point $x$, and therefore, can immediately be resolved.

  As a technical remark, in Case 2, $x$ might be the lower (resp. upper) bound[3] on the extent of $q$. If this happens, we simply discard $q_{left}$ (resp. $q_{right}$) as it would have a meaningless extent.

In either case, we are left with at most two queries (i.e., $q_{left}$ and $q_{right}$) to further process. The algorithm handles each of them recursively in the same manner.

**Example.** We use the dataset $D$ in Figure 3a to demonstrate the algorithm. Let $k = 4$. The first query is $q_1 = (-\infty, \infty)$. Suppose that the server responds by returning $R_1 = \{t_4, t_6, t_7, t_8\}$ and a signal that $q_1$ overflows. The $(k/2) = 2$-nd smallest tuple in $R_1$ is $t_6$ (after random tie breaking), whose value is $x = 55$. As $R_1$ has $c = 3$ tuples with value 55 and $c > k/4 = 1$, we perform a 3-way split on $q_1$ at 55, generating $q_2 = (-\infty, 54]$, $q_3 = [55, 55]$ and $q_4 = [56, \infty)$. As $q_3$ has degenerated into a point, it is resolved immediately, fetching $t_6, t_7$ and $t_8$. These tuples have already been extracted before, but this time they come with an extra fact that no more tuple can exist at point 55.

Let us look at $q_2$. Suppose that the server's response is $R_2 = \{t_1, t_2, t_4, t_5\}$, plus an overflow signal. Hence, $x = 20$ and $c = 1$. Thus, a two-way split on $q_2$ at 20 creates $q_5 = (-\infty, 19]$ and $q_6 = [20, 54]$. Queries $q_4$, $q_5$ and $q_6$ are all resolved.

**Analysis.** The lemma below bounds the cost of *rank-shrink*.

LEMMA 1. *When $d = 1$, rank-shrink requires $O(n/k)$ queries.*

---

[3] $x$ cannot be both because otherwise $q$ would be a point and therefore could not have overflow.

PROOF. The main tool used by our proof is a *recursion tree* $T$ that captures the spawning relationships of the queries performed by *rank-shrink*. Specifically, each node of $T$ represents a query. Node $u$ is the parent of node $u'$ if query $u'$ is created by a (2-way or 3-way) split of query $u$. Each internal node thus has 2 or 3 child nodes. Figure 3b shows the recursion tree for the queries performed in our earlier example on Figure 3a.

We focus on bounding the number of leaves in $T$ because it dominates the number of internal nodes. Observe that each leaf $v$ corresponds to a *disjoint* interval in $dom(A_1)$, due to the way splits are carried out. There are three types of $v$:

- Type-1: the query represented by $v$ is immediately resolved in a 3-way split (i.e., $q_{mid}$ in Case 2). The interval of $v$ contains at least $k/4$ (identical) tuples in $D$.
- Type-2: query $v$ is not type-1, but also covers at least $k/4$ tuples in $D$.
- Type-3: query $v$ covers less than $k/4$ tuples in $D$.

For example, among the leaf nodes in Figure 3, $q_3$ is of type 1, $q_5$ and $q_6$ are of type 2, and $q_4$ is of type 3.

As the intervals of various leaves cover disjoint bags of tuples, the number of type-1 and -2 leaves is at most $\frac{n}{k/4} = 4n/k$. Each leaf of type-3 must have a sibling in $T$ that is a type-2 leaf (i.e., in Figure 3, such a sibling of $q_4$ is $q_3$). On the other hand, a type-2 leaf has at most 2 siblings. It thus follows that there are at most twice as many type-3 leaves as type-2, i.e., the number of type-3 leaves is no more than $8n/k$. This completes the proof.

We remark that the above analysis implies that (quite loosely) $T$ has no more than $4n/k + 8n/k = 12n/k$ leaves. Thus, there cannot be more than this number of internal nodes in $T$. □

## 2.3 Rank-Shrink for Higher Dimensionality

We are now ready to extend *rank-shrink* to handle any $d > 1$. In addition to the ideas exhibited in the preceding subsection, we also apply an inductive approach: converting the $d$-dimensional problem to several $(d-1)$-dimensional ones. Our discussion below assumes that the $(d-1)$-dimensional problem has already been settled by *rank-shrink*.

Given a query $q$, the algorithm (as in 1d) sets out to solicit the server's response $R$, and finishes if $q$ is resolved. Otherwise, it examines whether $A_1$ is exhausted in $q$, i.e., whether the extent of $q$ on $A_1$ has only 1 value, say $x$, in $dom(A_1)$. If so, we can then focus on attributes $A_2, ..., A_d$. This is a $(d-1)$-dimensional version of Problem 1, in the $(d-1)$-dimensional subspace covered by the extents of $q$ on $A_2, ..., A_d$, eliminating $A_1$ by fixing it to $x$. Hence, we invoke *rank-shrink* to solve it.

Consider that $A_1$ is not exhausted on $q$. Similar to the 1d algorithm, we will split $q$ such that, either every resulting rectangle covers at least $k/4$ tuples in $R$, or one of them can be immediately solved as a $(d-1)$-dimensional problem. The splitting proceeds exactly as described in Cases 1 and 2 of Section 2. The only difference is that the rectangle $q_{mid}$ in Case 2 is not a point, but instead, a rectangle on which $A_1$ has been exhausted. Hence, $q_{mid}$ is processed as a $(d-1)$-dimensional problem with *rank-shrink*.

As with the 1d case, the algorithm recurses on $q_{left}$ and $q_{right}$ (provided that they have not been discarded for having a meaningless extent on $A_1$).

**Example.** We demonstrate the algorithm using the 2d dataset in Figure 4, where $D$ has 10 tuples $t_1, ..., t_{10}$. Let $k = 4$. The first query $q_1$ issued covers the entire data space. Suppose that the server responds with $R_1 = \{t_4, t_7, t_8, t_9\}$ and an overflow signal. We 3-way split $q_1$ at $A_1 = 80$ into $q_2$, $q_3$ and $q_4$, whose rectangles can be found in Figure 4. Specifically, the $A_1$-extents of $q_2, q_3, q_4$ are $(-\infty, 79], [80, 80], [81, \infty)$ respectively, while their $A_2$-extents are all $(-\infty, \infty)$. Note that $A_1$ is exhausted on $q_2$; alternatively, we can see that $q_2$ is equivalent to a 1d query on the vertical line $A_1 = 80$. Hence, $q_2$ is recursively settled by our 1d algorithm (which, as can be verified easily, requires 3 queries).

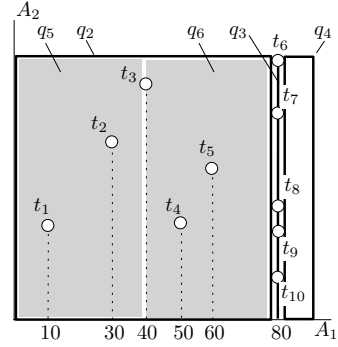

**Figure 4: Illustration of 2d *rank-shrink***

Suppose that the server's response to $q_2$ is $R_2 = \{t_2, t_3, t_4, t_5\}$ and an overflow signal. Accordingly, $q_2$ is split into $q_5$ and $q_6$ at $A_1 = 40$, whose rectangles are also shown in Figure 4. Finally, $q_4, q_5$ and $q_6$ are all resolved.

**Analysis.** We have the lemma below for general $d$:

LEMMA 2. *Rank-shrink performs $O(dn/k)$ queries.*

PROOF. The case $d = 1$ has been proved in Lemma 1. Next, assuming that *rank-shrink* issues at most $\alpha(d-1)n/k$ queries for solving a $(d-1)$-dimensional problem with $n$ tuples (where $\alpha$ is a positive constant), we will show that the cost is at most $\alpha dn/k$ for dimensionality $d$.

Again, our argument leverages a recursion tree $T$. As before, each node of $T$ is a query, such that node $u$ parents node $u'$, if query $u'$ was created from splitting $u$. We make a query $v$ a leaf of $T$ as soon as one of the following occurs:

- $v$ is resolved. We associate $v$ with a *weight* set to 1.
- $A_1$ is exhausted on rectangle $v$. Recall that such a query is solved as a $(d-1)$-dimensional problem. We associate $v$ with a weight, equal to the cost for *rank-shrink* for that problem.

For our earlier example with Figure 4, the recursion tree $T$ happens to be the same as the one in Figure 3b. The difference is that each leaf has a weight. Specifically, the weight of $q_3$ is 3 (i.e., the cost of solving the 1d query at the vertical line $A_1 = 80$ in Figure 4), and the weights of the other leaves are 1.

The total cost of *rank-shrink* on the $d$-dimensional problem, therefore, equals the total number of internal nodes in $T$, plus the total weight of all the leaves.

As the $A_1$-extents of the leaves' rectangles have no overlap, their rectangles cover disjoint tuples. Let us classify the leaves into types-1, -2 and -3 as in the proof of Lemma 1, by adapting the definition of type-1 in a straightforward fashion: $v$ is of this type if it is the middle node $q_{mid}$ from a 3-way split. Each type-3 leaf has weight 1 (as its corresponding query must be resolved). As proved in Lemma 1, the number of them is no more than $8n/k$.

Let $v_1, ..., v_\beta$ be all the type-1 and -2 nodes (i.e., suppose the number of them is $\beta$). Assume that node $v_i$ contains $n_i$ tuples of $D$. It holds that $\sum_{i=1}^{\beta} n_i \leq |D| = n$. The weight of $v_i$, by our inductive assumption, is at most $\alpha(d-1)n_i/k$. Hence, the total weight of all the type-1 and -2 nodes does not exceed $\alpha(d-1)n/k$.

The same argument in the proof of Lemma 1 shows that $T$ has less than $12n/k$ internal nodes. Thus, summarizing the above analysis, the cost of $d$-dimensional *rank-shrink* is no more than: $\frac{12n}{k} + \frac{8n}{k} + \alpha(d-1)\frac{n}{k} = (20 + \alpha(d-1))\frac{n}{k}$. To complete our inductive proof, we want $(20 + \alpha(d-1))\frac{n}{k}$ to be bounded from above by $\alpha dn/k$. This is true for any $\alpha \geq 20$. □

**Remark.** This concludes the proof of the first bullet of Theorem 1. We point out that when $d$ is fixed value (as is true in practice), the time complexity in Lemma 2 becomes $O(n/k)$, that is, asymptotically matching the trivial lower bound $n/k$. A natural question at this point is, if $d$ is not constant, is there an algorithm that can still guarantee cost $O(n/k)$? In Section 4, we will show that this is impossible.

## 3. CATEGORICAL ATTRIBUTES

We proceed to solve Problem 1 when the data space $\mathbb{D}$ is categorical. Recall that, as mentioned in Section 1.1, the domain $dom(A_i)$ of the $i$-th attribute $A_i$ is the set of integers in $[1, U_i]$ (where $U_i = |dom(A_i)|$), although it should be understood that the ordering of those integers is irrelevant. We will again first (in Section 3.1) clarify some preliminary concepts and give a baseline algorithm, before presenting the proposed solution (in Section 3.2).

### 3.1 Data Space Tree and Depth First Search

Unlike range predicates on numeric attributes, the predicate supported by the server on a categorical attribute $A_i$ ($1 \leq i \leq d$) is an equality constraint of the form $A_i = x$, where $x$ is either a value in $dom(A_i)$ or a wildcard $\star$. This difference prompts us to adopt an alternative approach to attack Problem 1. Instead of performing 2- or 3-way splits (as in the numeric case), we instead *enumerate* the points in $\mathbb{D}$. This idea looks drastic at first glance – $\mathbb{D}$ has a total of $\prod_{i=1}^d U_i$ points, where $U_i$ is the domain size of $A_i$. A naive way to enumerate the entire $\mathbb{D}$ is clearly intractable. It turns out, interestingly, that we can significantly reduce the cost from the formidable $\prod_{i=1}^d U_i$ to only roughly a linear term $\sum_{i=1}^d U_i$.

**Data space tree.** Let us start by arranging all the points of $\mathbb{D}$ into a tree $\mathbb{T}$, which we refer to as the *data space tree*. Each node $u$ of $\mathbb{T}$ represents a subspace enclosing all the points of $\mathbb{D}$ satisfying a *condition* like:

$$A_1 = c_1, ..., A_\ell = c_\ell, A_{\ell+1} = \star, ..., A_d = \star$$

where $\ell$ is an integer in $[0, d]$ and $c_1, ..., c_\ell$ are *not* wildcards. We say that $u$ is at *level* $\ell$. We will refer to the condition of $u$ as $query(u)$ because the condition obviously corresponds to a query that can be submitted to the server.

If we look at a point $(c_1, ..., c_d)$ in $\mathbb{D}$ as a string concatenating all its coordinates from the first to the $d$-th attribute, $\mathbb{T}$ can be regarded as a *trie* on all the $\prod_{i=1}^d U_i$ strings. Formally, the root of $\mathbb{D}$ is at level 0, and thus represents the entire $\mathbb{D}$ (equivalently, the condition of the root specifies a wildcard on all dimensions). In general, if $u$ is a level-$\ell$ node ($0 \leq \ell \leq d - 1$), it has $U_{\ell+1}$ child nodes of level $\ell + 1$. The $i$-th ($1 \leq i \leq U_{\ell+1}$) child $v$ of $u$ is such that, $query(v)$ agrees with $query(u)$ on all dimensions, except $A_{\ell+1}$ on which $query(v)$ specifies $A_{\ell+1} = i$. That is, $v$ refines $u$ on $A_{\ell+1}$ by setting this attribute to $i$. Each leaf of $\mathbb{T}$ is at level $d$ and represents a distinct point in $\mathbb{D}$.

To illustrate, Figure 5a shows a dataset $D$ with 10 tuples $t_1, ..., t_{10}$ in a 2d space $\mathbb{D}$ where each dimension has domain size 4. Figure 5b demonstrates the data space tree $\mathbb{T}$ (the subtrees of nodes $u_3$ and $u_4$ are omitted for simplicity). Node $u_1$, for instance, is associated with a $query(u_1)$ that has predicates $A_1 = \star$ and $A_2 = \star$, whereas $query(u_2)$ has predicates $A_1 = 1$ and $A_2 = \star$, and $query(u_7)$ has $A_1 = 1$ and $A_2 = 2$.

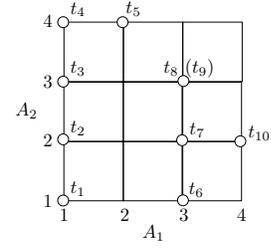

(a) A categorical dataset

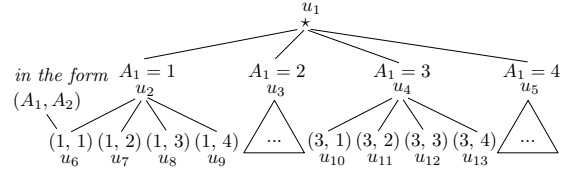

(b) The data space tree

**Figure 5: Illustration of categorical algorithms**

The following lemma presents a fact that will be useful later.

LEMMA 3. *Let $u, v$ be two nodes at the same level of $\mathbb{T}$. No tuple of $D$ can satisfy $query(u)$ and $query(v)$ at the same time.*

PROOF. Let the level be $\ell$. The conditions in $query(u)$ and $query(v)$ differ in their predicates on at least one attribute $A_i$ for some $i \in [1, \ell]$. No tuple can satisfy those predicates simultaneously. □

**Depth first search (DFS).** We now describe an algorithm, named *DFS*, that serves as the baseline approach. The algorithm simply traverses $\mathbb{T}$ in a depth-first manner. For each node $u$ in $\mathbb{T}$, it sends $query(u)$ to the server, and acquires the bag $R$ of tuples returned. As a pruning rule, if $query(u)$ is resolved, the subtree of $u$ no longer needs to be explored, as all the tuples in the subtree are already in $R$. If, on the other hand, $query(u)$ overflows, the algorithm processes each child of $u$ in the same manner.

Suppose $k = 3$. On the input of Figure 5a, *DFS* examines the nodes in the order $u_1, u_2, u_6, u_7, ...$ To see an example of pruning, consider the moment when *DFS* is at $u_3$. Since $query(u_3)$ is resolved (the query has predicates $A_1 = 2$ and $A_2 = \star$, and returns only $t_5$), the subtree of $u_3$ can thus be eliminated. It can be verified that *DFS* eventually visits all of $u_1, ..., u_{13}$.

**Remark.** Not surprisingly, *DFS* incurs expensive query cost in the worst case. We omit its analysis because it is tedious, and yet this algorithm is not the one advocated in this paper. In the next subsection, we give a better algorithm with the optimal performance.

### 3.2 Algorithm Slice-Cover

**Slice query.** We say that a query $q$ is a *slice query* if its predicates have the form:

$$..., A_{i-1} = \star, A_i = c, A_{i+1} = \star, ...$$

where $c$ is a value in $dom(A_i)$. Namely, the query has a wildcard predicate on *all but one* attribute $A_i$ for some $i \in [1, d]$. We use the notation $A_i = c$ to uniquely refer to a slice query. Clearly, varying $c$ in $dom(A_i)$ defines $U_i$ slice queries, such that the total number of slice queries of all dimensions is $\sum_{i=1}^d U_i$. In the example of Figure 5a, there are totally 8 slice queries.

**Slice-cover.** Now we describe an algorithm, named *slice-cover*, for solving Problem 1. The algorithm runs in two phases. In the

1117

| constant $c$ | 1 | 2 | 3 | 4 |
|---|---|---|---|---|
| $A_1 =$ | ovflow | $\{t_5\}$ | ovflow | $\{t_{10}\}$ |
| $A_2 =$ | $\{t_1, t_6\}$ | $\{t_2, t_7, t_{10}\}$ | $\{t_3, t_8, t_9\}$ | $\{t_4, t_5\}$ |

**Figure 6: Lookup table of slice queries ($k = 3$)**

*preprocessing phase*, we simply submit every slice query to the server, and record its response locally in a *lookup table* as follows. If a slice query $q$ is resolved, its result $q(D)$ (which has at most $k$ tuples) is entered in the table, whereas if $q$ overflows, we remember nothing but a bit indicating that $|q(D)| > k$. Figure 6 presents the contents of the lookup table for the example in Figure 5, assuming $k = 3$.

The second phase of *slice-cover* executes an algorithm, named *extended-DFS*, by supplying the root of the data space tree $\mathbb{T}$ as the parameter. In general, given a node $u$ in $\mathbb{T}$, *extended-DFS* returns all the tuples in $D$ satisfying $query(u)$ (hence, setting $u$ to the root of $\mathbb{T}$ settles Problem 1). *Extended-DFS* also performs a depth-first traversal of $\mathbb{T}$ (similar to the *DFS* algorithm of Section 3.1), but leverages the contents of the lookup table to boost the pruning effectiveness.

Without loss of generality, suppose that $u$ is at level $\ell$ (for some $\ell \in [0, d]$). *Extended-DFS* starts by sending $query(u)$ to the server. If $query(u)$ is resolved, the algorithm ends by reporting the result received from the server, because the subtree of $u$ can be eliminated (as explained in Section 3.1 for *DFS*). Next, we focus on the more interesting case where $query(u)$ overflows, implying that $u$ is an internal node of $\mathbb{T}$.

As with *DFS*, *extended-DFS* may access each child node $v$ of $u$. However, before doing so, it attempts to answer $query(v)$ *locally* using the lookup table (i.e., without bothering the server with another query). To explain, recall that $v$, which is at level $\ell + 1$, is associated with a $query(v)$ that refines $query(u)$. That is, $query(v)$ inherits the predicates of $query(u)$ on all attributes, except $A_{\ell+1}$ on which $query(v)$ has a predicate, say, $A_{\ell+1} = c$ for some $c \in dom(A_{\ell+1})$.

Let us observe that the result of $query(v)$ is *completely* contained in the result of the slice query $A_{\ell+1} = c$. Remember that the server's response, say $R$, to the slice query is already available in our lookup table. Therefore, we fetch $R$ from the table (at no cost), and see whether the slice query was resolved. If yes, $query(v)$ can be accurately answered by returning the tuples in $R$ satisfying $query(v)$, in which case the subtree of $v$ does not need to be explored further. If, on the other hand, $R$ shows that the slice query overflew, *extended-DFS* recursively processes $v$ in the same manner.

**Example.** We illustrate *extended-DFS* using the example of Figure 5. Let $k = 3$. The lookup table output by the preprocessing stage is in Figure 6.

The algorithm starts with $query(u_1)$, where $u_1$ is the root of $\mathbb{T}$ (Figure 5b). Even without sending $query(u_1)$ to the server, we know that it overflows for sure, because at least one slice query overflew in preprocessing, as is clear from the (lookup) table. Focusing on the first child $u_2$ of $u_1$, *extended-DFS* inspects the table to decide whether $query(u_2)$ can be answered locally. The inspection examines slice query $A_1 = 1$, i.e., the predicate by which $query(u_2)$ refines $query(u_1)$. The table indicates that the slice query overflew. Hence, we recursively apply *extended-DFS* on $u_2$.

At $u_2$, the algorithm checks the table as to whether $query(u_6)$ can be answered locally. This time, we focus on slice query $A_2 = 1$ (i.e., the extra predicate in $query(u_6)$ compared to $query(u_2)$), which turns out to be resolved. Hence, $query(u_6)$ is directly answered from the result $\{t_1, t_6\}$ of the slice query (i.e., returning only $t_1$, as $t_6$ does not qualify $query(u_6)$). Thus, *extended-DFS* does *not* recurse into $u_6$. Similarly, $query(u_7), query(u_8)$ and $query(u_9)$ can all be answered locally.

We now backtrack to the root of $u_1$, and turn attention to the second child $u_3$ of $u_1$. A lookup is carried out to see whether $query(u_3)$ can be acquired from the table. The answer is yes (using the result of slice query $A_1 = 2$). Hence, the subtree of $u_3$ is not explored further. The rest of the execution proceeds in the same manner. Overall, besides $u_1$, *extended-DFS* is also (recursively) invoked on $u_2$ and $u_4$. No query is ever issued to the server in the entire process.

**Heuristic.** Next, we give a heuristic that does not affect the worst-case cost of *slice-cover*, but can improve its performance on real data. The motivating rationale is that, some slice queries executed in the pre-processing phase may not eventually be needed. In any case, even if a slice query *does* need to be consulted by the algorithm, there is no harm to run the query at the *first* time such a need arises, and register the server's response in the lookup table. If the slice query is consulted for a second time, (as before) the query does not need to be re-issued, for its result is already available in the table. This allows us to get rid of the entire preprocessing phase. We refer to the algorithm equipped with this heuristic as *lazy-slice-cover*.

**Analysis.** We bound the performance of *slice-cover* with the following result, which also applies to *lazy-slice-cover* as it does not require any more query than *slice-cover*.

LEMMA 4. *If $d > 1$, slice-cover performs at most $\sum_{i=1}^{d} U_i + \frac{n}{k} \cdot \sum_{i=1}^{d} \min\{U_i, \frac{n}{k}\}$ queries. If $d = 1$, the number of queries is $U_1$.*

PROOF. For $d = 1$, *slice-cover* terminates right after the pre-processing phase, and hence, issues $U_1$ queries. For $d \geq 2$, the preprocessing phase obviously issues $\sum_{i=1}^{d} U_i$ queries. Next, we will show that *extended-DFS* incurs cost at most $\frac{n}{k} \cdot \sum_{i=1}^{d} \min\{U_i, \frac{n}{k}\}$.

Let $T$ be the nodes of $\mathbb{T}$ on which *extended-DFS* is invoked (i.e., $T$ includes the root of $\mathbb{T}$, and the nodes *extended-DFS* recurses into). For instance, on the example of Figure 5, our earlier discussion showed that $T$ includes nodes $u_1, u_2$ and $u_4$. The number of nodes in $T$ is an upper bound of the number of queries *extended-DFS* performs.

For each $i \in [1, d]$, let $S_i$ be the set of level-$i$ nodes in $T$. The following analysis will prove that $|S_i| \leq \frac{n}{k} \cdot \min\{U_i, \frac{n}{k}\}$, which is enough to complete the proof. To bound $|S_i|$, we consider $S_{i-1}$ (i.e., one level closer to the root). We will show:

- Fact 1: at most $n/k$ nodes in $S_{i-1}$ are internal in $T$.
- Fact 2: each internal node in $S_{i-1}$ can have at most $\min\{U_i, \frac{n}{k}\}$ child nodes in $T$.

Combining both facts gives the desired upper bound $\frac{n}{k} \cdot \min\{U_i, \frac{n}{k}\}$ on $|S_i|$.

*Proof of Fact 1.* For a node $u \in S_{i-1}$, denote by $n_u$ the number of tuples in $D$ satisfying $query(u)$. It follows from Lemma 3 that $\sum_{u \in S_{i-1}} n_u \leq |D| = n$.

On the other hand, as $u$ is an internal node in $T$, $n_u > k$. Otherwise, $query(u)$ would have been resolved, in which case the subtree of $u$ should have been pruned in *extended-DFS*, contradicting $u$ being an internal node. This, together with $\sum_{u \in S_{i-1}} n_u \leq n$, proves that $S_{i-1}$ has no more than $n/k$ internal nodes.

*Proof of Fact 2.* For each node $u \in S_{i-1}$, let $children(u)$ be the set of child nodes of $u$ in $T$. By the way $\mathbb{T}$ is defined, $|children(u)| = U_i$. As $u$ cannot have more child nodes in $T$ than in $\mathbb{T}$, Fact 2 holds if $U_i \leq n/k$. The rest of the proof assumes $U_i > n/k$.

For each node $v \in children(u)$, $query(v)$ refines $query(u)$ by replacing the predicate on $A_i$ (which is $A_i = \star$ in $query(u)$) with $A_i = c$ for some $c \in dom(U_i)$. Furthermore, each $v$ specifies a different $c$. Extended-DFS guarantees that $v$ should not be accessed (and hence, should not belong to $T$) if the slice query $A_i = c$ is resolved. In other words, if we denote by $Q$ the set of slice queries of the form $A_i = c$ (i.e., $|Q| = U_i$), the number of child nodes of $u$ in $T$ is at most the number of queries in $Q$ that *overflow*.

Clearly, no tuple of $D$ can satisfy both slice queries $A_i = c_1$ and $A_i = c_2$ as long as $c_1 \neq c_2$. In other words, the sum of $|q(D)|$ (i.e., the number of tuples satisfying $q$) of all $q \in Q$ is at most $n$. On the other hand, $|q(D)| \geq k$ if $q$ overflows. Thus, the number of overflowing queries in $Q$ is at most $n/k$. □

**Remark.** Lemma 4 establishes the second and third bullets of Theorem 1. As we will see in the next section, the upper bounds in the lemma cannot be improved by more than a constant factor in the worst case.

## 4. LOWER BOUND RESULTS

This section turns away from upper bounds, and focuses on the hardness of Problem 1. In Section 4.1 (4.2), we will give asymptotic lower bounds on how many queries are needed for settling the problem, when the data space $\mathbb{D}$ is numeric (categorical).

### 4.1 Numeric Attributes

The objective of this subsection is to establish:

THEOREM 3. *Let $k, d, m$ be arbitrary positive integers such that $d \leq k$. There is a dataset $D$ (in a numeric data space) with $n = m(k + d)$ tuples such that, any algorithm must use at least $dm$ queries to solve Problem 1 on $D$.*

It is, therefore, impossible to improve our algorithm *rank-shrink* (see Lemma 2) by more than a constant factor in the worst case, as shown below:

COROLLARY 1. *In a numeric data space, no algorithm can guarantee solving Problem 1 with $o(dn/k)$ queries.*

PROOF. If there existed such an algorithm, let us use it on the inputs in Theorem 3. The cost is $o(dn/k) = o(dm(k + d)/k)$ which, due to $d \leq k$, is $o(dm)$, causing a contradiction. □

We now proceed to prove Theorem 3, using a hard dataset $D$ as illustrated in Figure 7. The domain of each attribute is the set of integers from 1 to $m + 1$, namely, $\mathbb{D} = [1, m + 1]^d$. $D$ has $m$ groups of $d + k$ tuples. Specifically, the $i$-th ($1 \leq i \leq m$) group has $k$ tuples at the point $(i, ..., i)$, taking value $i$ on all attributes. We call them *diagonal tuples*. Furthermore, for each $j \in [1, d]$, Group $i$ also has a tuple that takes value $i + 1$ on attribute $A_j$, and $i$ on all other attributes. Such a tuple is referred to as a *non-diagonal* tuple. Overall, $D$ has $km$ diagonal and $dm$ non-diagonal tuples.

Let $S$ be the set of $dm$ points in $\mathbb{D}$ that are equivalent to the $dm$ non-diagonal tuples in $D$, respectively (i.e., each point in $S$ corresponds to a distinct non-diagonal tuple). As explained in Section 2.1, each query can be regarded as an axis-parallel rectangle in $\mathbb{D}$. With this correspondence in mind, we observe the following for any algorithm that correctly solves Problem 1 on $D$:

LEMMA 5. *When the algorithm terminates, each point in $S$ must be covered by a <u>distinct</u> resolved query already performed.*

|  | $A_1$ | $A_2$ | $\cdots$ | $A_d$ |  |
|---|---|---|---|---|---|
| Group 1 | 1 | 1 | $\cdots$ | 1 | $\}$ $k$ tuples |
|  | $\cdots$ |  |  |  |  |
|  | 1 | 1 | $\cdots$ | 1 |  |
|  | 2 | 1 | $\cdots$ | 1 |  |
|  | 1 | 2 | $\cdots$ | 1 | $\}$ $d$ tuples |
|  | $\cdots$ |  |  |  |  |
|  | 1 | 1 | $\cdots$ | 2 |  |
| Group 2 | 2 | 2 | $\cdots$ | 2 | $\}$ $k$ tuples |
|  | $\cdots$ |  |  |  |  |
|  | 2 | 2 | $\cdots$ | 2 |  |
|  | 3 | 2 | $\cdots$ | 2 |  |
|  | 2 | 3 | $\cdots$ | 2 | $\}$ $d$ tuples |
|  | $\cdots$ |  |  |  |  |
|  | 2 | 2 | $\cdots$ | 3 |  |
| $\vdots$ | $\cdots$ |  |  |  |  |
| Group $m$ | $m$ | $m$ | $\cdots$ | $m$ | $\}$ $k$ tuples |
|  | $\cdots$ |  |  |  |  |
|  | $m$ | $m$ | $\cdots$ | $m$ |  |
|  | $m+1$ | $m$ | $\cdots$ | $m$ |  |
|  | $m$ | $m+1$ | $\cdots$ | $m$ | $\}$ $d$ tuples |
|  | $\cdots$ |  |  |  |  |
|  | $m$ | $m$ | $\cdots$ | $m+1$ |  |

**Figure 7: A hard numeric dataset**

PROOF. Every point $p \in S$ must be covered by a resolved query. Otherwise, $p$ is either never covered by any query, or covered by only overflowing queries. In the former case, the tuple of $D$ at $p$ could not have been retrieved, whereas in the latter, the algorithm could not rule out the possibility that $D$ had more than one tuple at $p$. In neither case could the algorithm have terminated.

Next we show that no resolved query $q$ covers more than one point in $S$. Otherwise, assume that $q$ contains $p_1$ and $p_2$ in $S$, in which case $q$ fully encloses the minimum bounding rectangle, denoted as $r$, of $p_1$ and $p_2$. Without loss of generality, suppose that $p_1$ ($p_j$) is from Group $i$ ($j$) such that $i \leq j$. If $i = j$, then $r$ contains the point $(i, ..., i)$, in which case at least $k + 2$ tuples satisfy $q$ (i.e., $p_1, p_2$ and the $k$ diagonal tuples from Group $i$). Consider, on the other hand, $i < j$. In this scenario, the coordinate of $p_1$ is at most $i + 1 \leq j$ on all attributes, while the coordinate of $p_2$ is at least $j$ on all attributes. Thus, $r$ contains the point $(j, ..., j)$, causing at least $k + 2$ tuples to satisfy $q$ (i.e., $p_1, p_2$ and the $k$ diagonal tuples from Group $j$). Therefore, $q$ must overflow in any case, i.e., a contradiction. □

The lemma indicates that at least $|S| = dm$ queries must be performed, which validates the correctness of Theorem 3.

### 4.2 Categorical Attributes

First, if $d = 1$, $U_1$ is a trivial lower bound on the cost of solving Problem 1. The reason is that, as long as $D$ has more than $k$ tuples, we must issue a query $A_1 = c$ for every $c \in dom(A_1)$ to verify whether a tuple exists at point $c$. For $d > 1$, this subsection will establish:

THEOREM 4. *Let $k, d, U$ be positive integers satisfying $dU^2 \leq 2^{d/4}$, $U \geq 3$, $k \geq 3$, and $d = 2k$. There is a dataset $D$ with $n = dU$ tuples in a d-dimensional categorical space, where each attribute has a domain size $U$, such that, any algorithm must use $\Omega(dU^2)$ queries to solve Problem 1 on $D$.*

We thus know that our algorithm *slice-cover* (see Lemma 4) cannot be improved by more than a constant factor in the worst case, as shown below:

COROLLARY 2. *In a categorical data space, no algorithm can guarantee solving Problem 1 with $o(\sum_{i=1}^{d} U_i + \frac{n}{k} \sum_{i=1}^{d} \min\{U_i, \frac{n}{k}\})$ queries.*

|        | $A_1$ | $A_2$ | $\cdots$ | $A_d$ |
|--------|-------|-------|----------|-------|
| Group 0 $\begin{cases} \\ \\ \\ \\ \end{cases}$ | 1 0 $\vdots$ 0 | 0 1 $\vdots$ 0 | $\cdots$ $\cdots$ $\vdots$ $\cdots$ | 0 0 $\vdots$ 1 |
| Group 1 $\begin{cases} \\ \\ \\ \\ \end{cases}$ | 2 1 $\vdots$ 1 | 1 2 $\vdots$ 1 | $\cdots$ $\cdots$ $\vdots$ $\cdots$ | 1 1 $\vdots$ 2 |
| $\vdots$ | | | $\cdots$ | |
| Group $U-1$ $\begin{cases} \\ \\ \\ \\ \end{cases}$ | 0 $U-1$ $\vdots$ $U-1$ | $U-1$ 0 $\vdots$ $U-1$ | $\cdots$ $\cdots$ $\vdots$ $\cdots$ | $U-1$ $U-1$ $\vdots$ 0 |

**Figure 8: A hard categorical dataset**

PROOF. In the setting of Theorem 4, $\frac{n}{k} = \frac{dU}{k} = 2U$. Furthermore, $U_1 = ... = U_d = U$. Hence, $\frac{n}{k}\sum_{i=1}^{d}\min\{U_i, \frac{n}{k}\} = \frac{dU}{k} \cdot dU = 2dU^2$. Thus, the complexity in the corollary is $o(dU + dU^2) = o(dU^2)$. Hence, if the corollary was wrong, there would be an algorithm solving Problem 1 on the inputs in Theorem 4 with $o(dU^2)$ queries, which is a contradiction. □

The rest of the subsection serves as the proof of Theorem 3. Our discussion is based on a hard dataset $D$ illustrated in Figure 8. Formally, $D$ consists of $U$ groups, each of which has $d$ tuples. In the $i$-th ($0 \le i \le U-1$) group, for each attribute $A_j$ ($1 \le j \le d$), there is a tuple that takes value $(i+1) \mod U$ on $A_j$, and value $i$ on all the other $d-1$ attributes. In other words, the data space $\mathbb{D}$ is $[0, U-1]^d$, although readers should be reminded that using integers to represent the values of a (categorical) attribute is purely for convenience, and that the ordering of those integers is irrelevant.

As before, we say that a query *covers* a point $p \in \mathbb{D}$ if $p$ satisfies the query. The next lemma gives an important fact:

LEMMA 6. *If an algorithm has solved Problem 1 on our constructed $D$, then every point in $\mathbb{D}$ must be covered by at least one <u>resolved</u> query already performed.*

PROOF. We say that a point in $\mathbb{D}$ is *empty* if $D$ has no tuple at that point. Assume the existence of a point $p \in \mathbb{D}$ that is not covered in any resolved query. Hence, $p$ is either outside all the queries the algorithm issued, or is covered only by the overflowing queries. Hence, if $p$ was empty, the algorithm got no hint as to whether a tuple exists at $p$, and therefore, could not have terminated. On the other hand, if $p$ was not empty, the algorithm saw a tuple at $p$ but could not decide whether $D$ had any other tuple at $p$ (i.e., duplicates). In this case, the algorithm could not have terminated either. Thus we have a contradiction. □

We will show that any correct algorithm must perform $\Omega(dU^2)$ resolved queries. Recall that, on each attribute $A_i$ ($1 \le i \le d$), a query $q$ has either a *wildcard predicate* $A_i = \star$, or a *constant predicate* $A_i = c$ for some $c \in [0, U-1]$. We say that $q$ is *diverse*, if it has *at least* two non-wildcard predicates with different constants specified. For example, the query with predicates $A_1 = 1, A_2 = 2, A_3 = \star, ..., A_d = \star$ is diverse, and so is the query with $A_1 = 1, A_2 = 1, A_3 = 2, A_4 = \star, ..., A_d = \star$ (due to the predicates on $A_2$ and $A_3$), whereas the query with $A_1 = 1, A_2 = 1, A_3 = \star, ..., A_d = \star$ is not (as the same constant appears in the non-wildcard predicates).

LEMMA 7. *A diverse query $q$ has at most two qualifying tuples, and hence, is always resolved (since $k > 2$).*

PROOF. Let $c_1 \ne c_2$ be constants, each of which appears in a non-wildcard predicate of $q$. Suppose $c_1 < c_2$. Two tuples from different groups cannot satisfy $q$ simultaneously. Otherwise, assume that tuple $t_1$ from group $i$ and tuple $t_2$ from group $j$ qualify $q$, and that $i < j$ without loss of generality. As $t_1$ contains only $i$ and $i+1$ in its attributes, we know $c_1 = i$ and $c_2 = i+1$. If $j < U-1$, $t_2$ contains only $j$ and $j+1$ in its attributes. We thus require $c_1 = j = i$, which violates $i < j$. If $j = U-1$, $t_2$ contains only 0 and $U-1$ in its attributes. In this case, $c_1 = 0 = i$ and $c_2 = U-1 = i+1$, which is also impossible because $U > 2$.

On the other hand, it is easy to see that, no three tuples from the same group can together take value $c_1$ on one attribute, and also, value $c_2$ on another attribute. The lemma then follows. □

We say that a query $q$ is *monotonic*, if (i) $q$ has *at least* two non-wildcard predicates, and (ii) the same constant is specified in *all* the non-wildcard predicates. For example, the query with predicates $A_1 = 1, A_2 = 1, A_3 = \star, ..., A_d = \star$ is monotonic, whereas the query with $A_1 = 1, A_2 = 2, A_3 = \star, ..., A_d = \star$ is not (as different constants are used in the non-wildcard predicates), and neither is the query with $A_1 = 1, A_2 = \star, ..., A_d = \star$ (as it has only one non-wildcard predicate).

LEMMA 8. *A resolved monotonic query $q$ has at least $d/2$ non-wildcard predicates, and hence, covers at most $2^{d/2}$ points in $\mathbb{D}$.*

PROOF. Let $c$ be the constant in all the non-wildcard predicates of $q$. If $q$ has $\lambda \ge 2$ non-wildcard predicates, it retrieves exactly $d - \lambda$ tuples from group $c$, and no tuple from any other group. Hence, for $q$ to be resolved, $d - \lambda$ cannot exceed $k$, that is, $\lambda \ge d - k = d/2$. □

It turns out that if a query is resolved, it *must* be either diverse or monotonic. In fact, if a query $q$ is neither diverse nor monotonic, it has at most one non-wildcard predicate. Such $q$ must retrieve at least $d$ tuples and hence, overflow (recall that $d = 2k > k$).

Given two different integers $x, y$ in $[0, U-1]$, we define a *bichromatic set* $S(x, y)$ of points in $\mathbb{D}$:

> for each $i \in [1, d]$, $S(x, y)$ includes all the points that take $x$ or $y$ as their values on attribute $A_i$, except points $(x, x, ..., x)$ and $(y, y, ..., y)$.

For example, for $d = 3$ and $U = 3$, $S(1, 2) = \{(1, 1, 2), (1, 2, 1), (1, 2, 2), (2, 1, 1), (2, 1, 2), (2, 2, 1)\}$. That is, $S(1, 2)$ has all the points having only 1 or 2 as their coordinates, but does *not* contain $(1, 1, 1)$ and $(2, 2, 2)$. Clearly, there are $\binom{U}{2}$ bichromatic sets, each of which has $2^d - 2$ points.

We are ready to explain why $\Omega(dU^2)$ queries are necessary to settle Problem 1 on $D$. Our discussion considers only the situation where less than $\frac{d}{8}\binom{U}{2}$ diverse queries are performed by the algorithm (otherwise, trivially there are $\frac{d}{8}\binom{U}{2} = \Omega(dU^2)$ queries). By Lemma 6, every point of each bichromatic set must be covered by some resolved query. If a query $q$ covers at least one point in a bichromatic set $S(x, y)$, we say that $q$ *touches* $S(x, y)$.

A diverse query can touch at most one bichromatic set. As there are less than $\frac{d}{8}\binom{U}{2}$ diverse queries but $\binom{U}{2}$ bichromatic sets, we can find a bichromatic set that is touched by less than $d/8$ diverse queries. Let $S(\alpha, \beta)$ be that bichromatic set (for some $\alpha, \beta$ in $[0, U-1]$), and $Q$ the set of diverse queries touching it (thus, $|Q| < d/8$).

Consider any query $q \in Q$. Since $q$ touches $S(\alpha, \beta)$, $q$ has two non-wildcard predicates $A_i = \alpha$ and $A_j = \beta$ for some $i, j \in [1, d]$ with $i \ne j$ (in case multiple pairs of $(i, j)$ satisfy this requirement, choose one arbitrarily). Refer to $A_i$ and $A_j$ as the *salient attributes* of $q$. It is clear that $q$ cannot cover those points $p \in S(\alpha, \beta)$ such that $p$ has value $\alpha$ on *both* salient attributes of $q$. Let $salient(Q)$ be the union of the salient attributes of all the queries in $Q$. Hence, $|salient(Q)| < d/4$. By the previous reasoning, if a point $p \in S(\alpha, \beta)$ takes value $\alpha$ on *all* the attributes in $salient(Q)$, $p$ is not covered by *any* of the queries in $Q$. How many such $p$ are there? As $|salient(Q)| < d/4$, $p$ can still choose $\alpha$ or $\beta$ as its value on each of the at least $\frac{3}{4}d+1$ attributes *outside* $salient(Q)$. Excluding points $(\alpha, ..., \alpha)$ and $(\beta, ..., \beta)$, we know that the number of such $p$ is at least $2^{1+3d/4} - 2 \geq 2^{3d/4}$. As a remark, although we required $p$ to take $\alpha$ on the attributes in $salient(Q)$, the argument works as well by taking $\beta$.

We have shown that at least $2^{3d/4}$ points in $S(\alpha, \beta)$ have not been covered by diverse queries. Those points must be covered by resolved monotonic queries, each of which, by Lemma 8, contains at most $2^{d/2}$ points in $\mathbb{D}$. Therefore, the number of such queries must be at least $2^{3d/4}/2^{d/2} = 2^{d/4}$ which, as stated in Theorem 4, is at least $dU^2$, thus completing the proof.

## 5. EXTENSIONS: MIXED ATTRIBUTES

This section will extend our techniques to handle a mixed data space $\mathbb{D}$ that has both numeric and categorical attributes. As defined in Section 1.1, we consider without loss of generality that the first $cat$ attributes $A_1, ..., A_{cat}$ are categorical, and the other $d - cat$ attributes $A_{cat+1}, ..., A_d$ are numeric.

Define $\mathbb{D}_{\text{CAT}}$ as the categorical data (sub) space $dom(A_1) \times ... \times dom(A_{cat})$, namely, involving all and only the categorical attributes. Put differently, for any point $p \in D$, trimming off its coordinates on $A_{cat+1}, ..., A_d$ gives a $cat$-dimensional point $p_{\text{CAT}}$ in $\mathbb{D}_{\text{CAT}}$. In a natural manner, $p_{\text{CAT}}$ determines a set of points, which we denote by $\mathbb{D}_{\text{NUM}}(p_{\text{CAT}})$, defined as:

> $\mathbb{D}_{\text{NUM}}(p_{\text{CAT}})$ is the set points $p' \in \mathbb{D}$, such that $p'$ shares the same value as $p_{\text{CAT}}$ on every categorical attribute.

In fact, $\mathbb{D}_{\text{NUM}}(p_{\text{CAT}})$ decides a $(d - cat)$-dimensional numeric subspace of $\mathbb{D}$: it includes all the numeric dimensions of $\mathbb{D}$, while fixing the categorical attributes to those of $p_{\text{CAT}}$.

As an example, the scenario of Figure 1 has four attributes, among which $cat = 2$ are categorical: $A_1 =$ MAKE and $A_2 =$ BODY STYLE. Let $p_{\text{CAT}} =$ (BMW, sedan). Then, $\mathbb{D}_{\text{NUM}}(p_{\text{CAT}})$ includes all the points $p' \in \mathbb{D}$ whose MAKE and BODY STYLE are BMW and sedan, respectively (but no constraint is imposed on the values of $p'$ along the numeric attributes $A_3 =$ PRICE and $A_4 =$ MILEAGE).

**Hybrid.** Next, we present an algorithm, named *hybrid*, to solve Problem 1 in a mixed $\mathbb{D}$. The algorithm combines *lazy-slice-cover* (see Section 3.2) and *rank-shrink* (Section 2.3). Roughly speaking, it first enumerates all the points in $\mathbb{D}_{\text{CAT}}$ using *lazy-slice-cover* and, when a point $p_{\text{CAT}} \in \mathbb{D}_{\text{CAT}}$ has been reached, invokes *rank-shrink* in the numeric subspace determined by $\mathbb{D}_{\text{NUM}}(p_{\text{CAT}})$.

Now we provide the missing details. Recall that, *lazy-slice-cover* runs on a *categorical server* (i.e., one that supports only categorical attributes). To apply it on a server of our context here, we set a query's predicate on numeric attribute $A_i$ (for each $i \in [cat + 1, d]$) to $A_i = (-\infty, \infty)$. The effect is to disregard all the numeric attributes, and hence, essentially emulates a categorical server.

Also recall that *lazy-slice-cover* performs an (improved) depth-first traversal on the data space tree $\mathbb{T}_{\text{CAT}}$ built from $\mathbb{D}_{\text{CAT}}$. Consider that it has come to a leaf node of $\mathbb{T}_{\text{CAT}}$, or equivalently, a point $p_{\text{CAT}} \in \mathbb{D}_{\text{CAT}}$. In Section 3.2, the processing of $p$ finished with one extra query, but *hybrid* invokes *rank-shrink* upon $\mathbb{D}_{\text{NUM}}(p_{\text{CAT}})$. Remember, however, that *rank-shrink* operates on a *numeric server* (i.e., one that supports only numeric attributes). To apply it on $\mathbb{D}_{\text{NUM}}(p_{\text{CAT}})$, we fix a query's predicate on categorical attribute $A_i$ (for each $i \in [1, cat]$) to $A_i = c_i$, where $c_i$ is the $A_i$ value of $p_{\text{CAT}}$. This effectively emulates a numeric server over the $(d-cat)$-dimensional numeric subspace implied by $\mathbb{D}_{\text{NUM}}(p_{\text{CAT}})$.

**Upper bounds.** Denote by $U_i$ ($1 \leq i \leq cat$) the domain size of the $i$-th categorical attribute $A_i$. The following lemma gives the performance guarantees of *hybrid*, and establishes the last two bullets of Theorem 1.

LEMMA 9. *When $cat > 1$, hybrid performs $\frac{n}{k} \sum_{i=1}^{cat} \min\{U_i, \frac{n}{k}\} + \sum_{i=1}^{cat} U_i + O((d-cat)n/k)$ queries. When $cat = 1$, the number of queries is $U_1 + O(dn/k)$.*

PROOF. We focus on only $cat > 1$ because the same argument also applies to $cat = 1$. For $cat > 1$, the term $\frac{n}{k} \sum_{i=1}^{cat} \min\{U_i, \frac{n}{k}\} + \sum_{i=1}^{cat} U_i$ is the cost of running *lazy-slice-cover*, and follows directly from Lemma 4.

Let $S$ be all the leaf nodes of $\mathbb{T}_{\text{CAT}}$ accessed by *hybrid*, namely, an instance of *rank-shrink* was executed on each node $u \in S$. Let $n_u$ be the number of tuples in $D$ having the same value as $u$ on every categorical attribute – these are the tuples in $\mathbb{D}_{\text{NUM}}(p_{\text{CAT}})$ where $p_{\text{CAT}}$ is the point in $\mathbb{D}_{\text{CAT}}$ corresponding to $u$. As disjoint bags of tuples are counted by the $n_u$ of different $u$, we have $\sum_{u \in S} n_u \leq n$. By Lemma 2, the instance of *rank-shrink* at $u$ issues $O((d-cat)n_u/k)$ queries. Hence, the total number of queries issued by *rank-shrink* is $O((d-cat) \sum_{u \in S} (n_u/k)) = O((d-cat)n/k)$. □

**Lower bounds.** We conclude this section by explaining why it is not possible to improve the upper bounds in Lemma 9 by more than a constant factor in the worst case. In fact, since a mixed data space is more general than both categorical and numeric spaces, the lower bounds in Section 4.1 and 4.2 are still applicable here. For example, if $cat = 0$, a mixed $\mathbb{D}$ becomes numeric. Hence, if there was an algorithm guaranteed faster than *hybrid* by non-constant times, that algorithm would terminate with $o(dm)$ queries on the inputs stated in Theorem 3, giving a contradiction. On the other hand, if $cat = d$, a mixed $\mathbb{D}$ becomes categorical. Similarly, if there was an algorithm faster than *hybrid* by non-constant times, that algorithm would terminate with $o(dU^2)$ queries on the inputs in Theorem 4, again giving a contradiction. The above discussion has taken care of $cat > 1$ or $cat = 0$, but an analogous argument apparently applies to the remaining case $cat = 1$ as well. This, together wish Theorems 3 and 4, establish Theorem 2.

## 6. EXPERIMENTS

In this section, we empirically evaluate the proposed techniques, and establish their superiority over alternative solutions.

**Data.** Our experiments were based on three real datasets, whose attributes, as well as the domain size of each attribute, are given in Figure 9 (where *num* indicates a numeric attribute). Specifically:

- *Yahoo* contains 69,768 tuples in a hidden database at *autos.yahoo.com*. Each tuple depicts 6 attributes of a vehicle. This is a mixed dataset (due to the presence of both numeric and categorical attributes).

- *NSF* contains 47,816 tuples in a hidden database at *nsf.gov/awardsearch*. Each tuple has 9 attributes of an NSF award. This is a categorical dataset.

- *Adult* contains 45,222 tuples in a census dataset that can be downloaded from *archive.ics.uci.edu/ml/datasets/adult*. Each tuple describes 14 attributes of a person working in the US. As with *Yahoo*, this is also a mixed dataset.

Figure 9: Attributes and their domain sizes of the datasets deployed

| Yahoo | | | | | | NSF | | | | | | | |
|---|---|---|---|---|---|---|---|---|---|---|---|---|---|
| OWNER | BODY-STYLE | MAKE | MILEAGE | YEAR | PRICE | AMNT | INSTRU | FIELD | PI-STATE | NSF-ORG | PROG-MGR | CITY | PI-ORG | PI-NAME |
| 2 | 7 | 85 | num | num | num | 5 | 8 | 49 | 58 | 58 | 654 | 1093 | 3110 | 29042 |

| Adult | | | | | | | | | | | | | |
|---|---|---|---|---|---|---|---|---|---|---|---|---|---|
| SEX | RACE | REL | EDU | MARITAL | WRK-CLASS | OCC | COUNTRY | EDU-NUM | AGE | WRK-HR | CAP-LOSS | CAP-GAIN | FNALWGT |
| 2 | 5 | 6 | 6 | 7 | 8 | 14 | 41 | num | num | num | num | num | num |

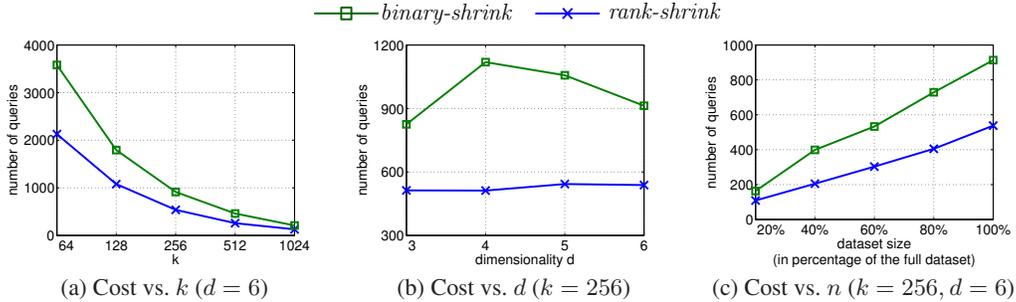

Figure 10: Query cost of numeric algorithms (dataset *Adult-numeric*)

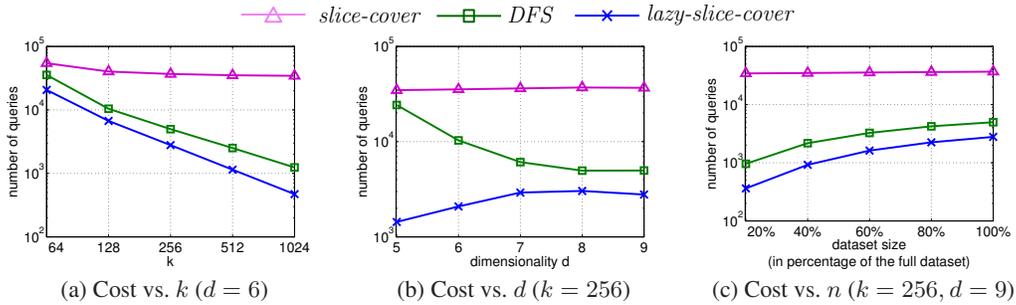

Figure 11: Query cost of categorical algorithms (dataset *NSF*)

We also extracted a numeric dataset from *Adult*, by including only its numeric attributes. The resulting dataset, named *Adult-numeric*, therefore has the same cardinality and dimensionality as *Adult*.

Recall that every algorithm in Sections 2, 3 and 5 works with an ordering of the attributes in the underlying dataset (i.e., which attribute is $A_1$, which one is $A_2$, and so on). The ordering is exactly as shown in Figure 9 (from left to right in each table), and is the same for all algorithms. Since our experiments require adjusting the value of $k$, we implemented a local server to run our algorithms. Our implementation conforms strictly to the problem setup in Section 1.1, so that the cost reported would be equivalent if the algorithms were executed on a remote web server. In a dataset, each tuple is assigned a random priority, so that if a query overflows, always the $k$ tuples with the highest priorities are returned.

**Numeric algorithms.** We start by studying the performance of numeric algorithms *binary-shrink* and *rank-shrink* in Section 2, using dataset *Adult-numeric*. For each algorithm, Figure 10a shows the number of queries it issued to extract *Adult-numeric* as a function of $k$. Setting $k$ to the median value 256, Figure 10b plots their efficiency as the dimensionality $d$ varies from 3 to 6. In this experiment, for each $d \in [3, 6]$, we created a $d$-dimensional dataset by taking the $d$ attributes of *Adult-numeric* that have the highest numbers of distinct values. Specifically, the attribute with the most distinct values is FNALWGT, the second is CAP-GAIN, followed by CAP-LOSS, WRK-HR, AGE and EDU-NUM. Using $k = 256$ and fixing $d$ to its original value 6, Figure 10c compares the two algorithms with respect to the dataset size $n$. Here, a 20% dataset corresponds to a random sample set of *Adult-numeric*, by independently sampling each of its tuples with a 20% probability. The datasets of the other percentages were generated in the same fashion.

*Rank-shrink* consistently outperformed *binary-shrink* in all cases. Furthermore, as predicted by our analysis in Section 2.3 (particularly, Lemma 2), the cost of *rank-shrink* was linear to $n$ and inversely linear to $k$ (to observe the inverse linearity, notice from Figure 10a that *rank-shrink* entailed half as many queries each time $k$ doubled). As a pleasant surprise, Figure 10b demonstrates that the cost of *rank-shrink* stayed nearly the same as $d$ increased, even though Lemma 2 indicates that the cost should grow linearly in the worst case. In fact, by looking at the proof of Lemma 2, one would realize that the presence of $d$ in the final time complexity is due to 3-way splits (see Figure 2). Such a split happens only if many tuples share the same value on a certain attribute. As this is not true in *Adult-numeric*, 3-way splits were seldom performed, which explains the phenomenon in Figure 10b.

**Categorical algorithms.** We deployed a similar methodology to compare the efficiency of algorithms *DFS*, *slice-cover* and *lazy-slice-cover* in Section 3. Note that *DFS* is in fact the baseline approach for crawling which was outlined in [15]. The underlying dataset was *NSF*. Figure 11 presents the results for the same set of experiments as in Figure 10. It is worth pointing out that, in Figure 11b, a $d$-dimensional dataset ($d \in [5, 9]$) was generated from *NSF* in the way as mentioned earlier for Figure 10b (the number of distinct values on each attribute equals the attribute's domain size, which can be found in Figure 9).

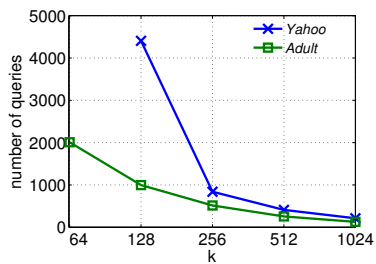

Figure 12: Cost of the mixed algorithm *hybrid*

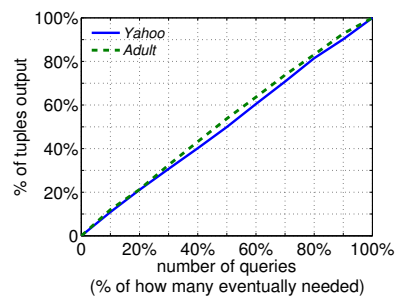

Figure 13: Output progressiveness of *hybrid* ($k = 256$)

Interestingly, *slice-cover*, even being asymptotically optimal, turned out to exhibit the worst performance. Of course, this does not contradict our theoretical analysis, because the optimality of *slice-cover* is reflected on the *hardest* dataset (see, for example, Figure 8). *NSF* is not such a dataset; hence, *slice-cover* does not guarantee better efficiency than a suboptimal solution like *DFS*. It is important to note that *lazy-slice-cover* was the clear winner in all the experiments (notice that the y-axes of all diagrams in Figure 11 are in log scale). The huge improvement of *lazy-slice-cover* over *slice-cover* confirms the necessity and effectiveness of the heuristic discussed in Section 3.2.

**Hybrid algorithms.** Having demonstrated the superiority of *rank-shrink* and *lazy-slice-cover* over their competitors, next we evaluate the behavior of their combination: the *hybrid* algorithm in Section 5. For this purpose, we employed both mixed datasets *Yahoo* and *Adult*. Figure 12 illustrates the number of queries performed by *hybrid* to crawl each dataset entirely, as $k$ changes from 64 to 1024. Note that there is no reported value for *Yahoo* at $k = 64$ because it has more than 64 identical tuples (i.e., they agree with each other on every dimension) – for the reason explained in Section 1.1, no algorithm can successfully extract the dataset in full when $k = 64$.

In practice, it would be a nice property for a crawling algorithm to be able to return the tuples of a hidden database in a *progressive* manner. Namely, it should gradually churn out new tuples as it runs, instead of outputting most tuples only at the end. This property allows the crawler to terminate the algorithm at any moment, while still able to obtain a number of tuples proportional to the amount of time that has been spent. Motivated by this, the last set of experiments examines the progressiveness of *hybrid*. In Figure 13 (where $k$ was set to 256), for each dataset, we present the percentage of the tuples extracted (the y-axis) against the percentage of the queries issued (the x-axis). For example, a point (20%, 30%) in this figure means that, *hybrid* is able to discover 30% of the tuples in the dataset, at the moment when it has issued 20% of all the queries that eventually need to be performed. We were delighted to observe linear progressiveness for both datasets, as is clear from Figure 13.

## 7. CONCLUSIONS

Currently, search engines cannot effectively index hidden databases, and are thus unable to direct queries to the relevant data in those repositories. With the rapid growth in the amount of such hidden data, this problem has severely limited the scope of information accessible to ordinary Internet users. In this paper, we attacked an issue that lies at the heart of the problem, namely, how to crawl a hidden database in its entirety with the smallest cost. We have developed algorithms for solving the problem when the underlying dataset has only numeric attributes, only categorical attributes, or both. All our algorithms are asymptotically optimal, i.e., none of them can be improved by more than constant times in the worst case. Our theoretical analysis has also revealed the factors that determine the hardness of the problem, as well as how much influence each of those factors has on the hardness.